# Rapid sampling through quantum computing*

Lov K. Grover

1D-435 Bell Labs, 700 Mountain Avenue, Murray Hill NJ 07974

lkgrover@bell-labs.com

## Abstract

This paper extends the quantum search class of algorithms to the multiple solution case. It is shown that, like the basic search algorithm, these too can be represented as a rotation in an appropriately defined two dimensional vector space. This yields new applications - an algorithm is presented that can create an arbitrarily specified quantum superposition on a space of size $N$ in $O(\sqrt{N})$ steps. By making a measurement on this superposition, it is possible to obtain a sample according to an arbitrarily specified classical probability distribution in $O(\sqrt{N})$ steps. A classical algorithm would need $\Omega(N)$ steps.

------------------------------------------------------------------

* This research was supported in part by funding from the U. S. Army Research Office under contract no. DAAG55-98-C-0040.

## 1. Introduction

*(i) Sampling:* Sampling is a fundamental technique for approximating answers that cannot be directly, or efficiently, computed. The uses of sampling for gathering information is the primary domain of the field of statistics. Sampling has also been linked to complexity theory, e.g counting, volume estimation, primality testing. For some applications uniform sampling is enough but for most applications one needs sampling according to a specified probability distribution. It is known how to rapidly generate samples when the probability distribution has a certain structure, e.g. when the distribution is log concave, samples can be generated in logarithmic time [1]. In general, the probability distributions are only specified implicitly, i.e. it is possible to calculate the probability of any given point easily but there is no known structure to the distribution. It is not possible to sample precisely according to an arbitrary probability distribution in fewer than $\Omega(N)$ classical steps. For example, it is easily seen that we will need to examine at least half the points, for if we leave out half of the points we could be missing a point with an arbitrarily high probability, e.g. the situation where all the probability is localized in a single point.

A quantum mechanical system is in multiple states simultaneously, quantum computing algorithms such as quantum search make use of this feature. This paper extends quantum search to develop a quantum algorithm that can create an arbitrarily specified super-position (and hence an arbitrary probability distribution) on $N$ points in only $O(\sqrt{N})$ steps.

*(ii) Quantum computing:* Just as classical digital systems can be constructed out of two state systems called bits, quantum mechanical systems can be constructed out of basic two state quantum mechanical systems called *qubits*. Quantum mechanical operations that can be carried out in a controlled way are unitary operations that act on a small number of qubits in each step. A good starting point to think of quantum mechanical algorithms is probabilistic algorithms [2] (e.g. simulated annealing). In a quantum mechanical algorithm, the system is started in a state that is easy to prepare on which a sequence of simple operations is applied. When the system is observed after applying these operations, it gives the answer to a difficult computational problem with a high probability.

Just like classical probabilistic algorithms, quantum mechanical algorithms work with a probability distribution over various states. However, unlike classical systems, the probability vector does not completely describe the system. In order to completely describe the system we need the *amplitude* in each state which is a complex number - such a specification is called an amplitude vector or a superposition. The probabilities in any state are given by the square of the absolute values of the amplitude in that state. The evolution of the system is obtained by premultiplying this amplitude vector by a state transition matrix, the entries of which are complex in general. It can be shown that in order to conserve probabilities, the state transition matrix has to be unitary, i.e. the columns of the transformation matrix are orthonormal [2]. It is easily shown that the following three transformations are unitary:

(i) NOT - a one-input one-output gate. The output is the inversion of the input.

(ii) CNTRL-NOT - a two-input two-output gate. The first output is the same as the first input. If the first input is 1, the second output is the inversion of the second input; if the first input is 0, the second output is equal to the second input.

(iii) CNTRL-CNTRL-NOT - a three-input three-output gate. The first two outputs are equal to the first two inputs respectively. If the first two inputs are both 1s, the third output is the inversion of the third input; if either of the first two inputs is 0, the third output is equal to the third input.

These are the analogs of classical NAND and NOR gates. Using these three gates it is possible to synthesize any boolean function $f(\bar{x})$ that can be synthesized classically with approximately the same number of gates. In order to develop more powerful quantum mechanical algorithms, we need some operations that are basically quantum mechanical, i.e. the entries in the state transition matrix are not all 0's and 1's. The two such operations that we will need in quantum search are the Walsh Hadamard (W-H) transformation and the selective inversion of the phase of certain states, these are discussed in the following two paragraphs.

A basic operation in quantum computing is the operation $M$ performed on a single qubit - this is represented by the following matrix: $M \equiv \frac{1}{\sqrt{2}} \begin{bmatrix} 1 & 1 \\ 1 & -1 \end{bmatrix}$, i.e. the state 0 is transformed into a superposition where the two states 0 & 1 have the same amplitude of $\frac{1}{\sqrt{2}}$, this superposition is denoted by: $\frac{1}{\sqrt{2}}(|0\rangle + |1\rangle)$; similarly state 1 is transformed into the superposition

$\frac{1}{\sqrt{2}}(|0\rangle - |1\rangle)$. A system consisting of $n$ qubits has $N \equiv 2^n$ basis states. If we perform the transformation $M$ on each qubit independently in sequence, the state transition matrix representing this operation is of dimension $N \times N$. Consider a case when the starting state is one of the $N$ basis states, i.e. a state described by a general string of $n$ binary digits composed of some 0s and some 1s. The result of performing the transformation $M$ on each qubit will be a superposition of states consisting of all possible $n$ bit binary strings with amplitude of each state being $\pm 2^{-\frac{n}{2}}$, i.e. $\pm\frac{1}{\sqrt{N}}$. This transformation is referred to as the W-H transformation [8].

The other transformation that we need is the selective phase inversion of the amplitude in certain states. The transformation matrix describing this for a 4 state system with selective phase inversion of the second state is: $\begin{bmatrix} 1 & 0 & 0 & 0 \\ 0 & -1 & 0 & 0 \\ 0 & 0 & 1 & 0 \\ 0 & 0 & 0 & 1 \end{bmatrix}$. Unlike the W-H transformation, the probability in each state stays the same. A realization of this kind of transformation can be achieved using the gates discussed so far [4].

In the exhaustive search problem, a function $f(\bar{x})$ defined over $N$ states (denoted by $\bar{x}$) is given which is known to be non-zero at a single value of $\bar{x}$, say $t$ ($t$ for target) - the goal is to find $t$. If there was no other information about $f(\bar{x})$ and one were using a classical computer, then on the average it would take $\frac{N}{2}$ function evaluations to solve this problem successfully. [9] found a quantum mechanical algorithm that took only $O(\sqrt{N})$ steps.

The quantum search algorithm [9] consisted of $\sqrt{N}$ repetitions of the operator $-I_t W I_{\bar{0}} W$ starting with the state $\bar{0}$ (here $W$ denotes the W-H transformation, $I_t$ denotes the selective phase inversion of the target state $t$, $I_{\bar{0}}$ denotes the selective phase inversion of the $\bar{0}$ state). Later on it was discovered that similar results are obtained by replacing the W-H transform by almost any quantum mechanical operation (say $U$) and the state $\bar{0}$ by any basis state $s$. It was shown that by starting with the basis state $s$, and carrying out $O\left(\frac{1}{|U_{ts}|}\right)$ repetitions of the operation sequence $-I_s U^{-1} I_t U$, one could reach the $t$ state [10] (similar results are also proved in [5]). This showed that one could use any starting state and unitary operation $U$ and from these amplify the amplitude in a desired target state $t$. A new class of algorithms was thus invented. These extended far beyond search problems - in fact, it was shown that this framework could be used to enhance almost any quantum mechanical algorithm [10].

One constraint with these algorithms was that they worked when the problem has exactly one $t$ state. For many problems like game-tree search, this was a major restriction and the algorithms either could not be shown to work with multiple solutions or needed complicated workarounds [6]. The paper [11] mentioned the multi-solution case as an open problem.

**2. This paper** This paper shows how the quantum search algorithm with general unitary transformations can be extended to the multi-solution case. From this, an algorithm for generating an arbitrarily specified superposition is developed. This can be used to sample according to a general probability distribution - the number of steps required is approximately the square-root of that of the corresponding classical algorithm. The following is the organization of the rest of this paper:

- section 3 carries out an analysis of the generalized search algorithm with an arbitrary number of solutions,
- section 4 considers the case with a single solution,
- section 5 extends this to the multi-solution case,
- section 6 shows how this can be used to generate an arbitrarily specified superposition.

## 3. Generalized search (arbitrary number of solutions)

The algorithm starts with the system in an $s$ state ($s$ for source), the object is to drive the system into the $t$ states, note that there is a single $s$ state but multiple $t$ states. The following sections show how this can be accomplished by means of an operation $Q$ which is defined as the composite operation $Q \equiv -I_s U^{-1} I_t U$ (note that, following standard matrix notation, this denotes the following sequence of operations: first $U$, then $I_t$, then $U^{-1}$ and finally $-I_s$.)

$I_s$ is a diagonal matrix with all diagonal elements equal to 1 except the $ss^{th}$ elements which is $-1$. In Dirac notation, which is commonly used for matrix operations in quantum mechanics, $I_s$ may be written as $I_s = I - 2|s\rangle\langle s|$. Here $I$ is the identity matrix and $|s\rangle$ the column vector with all except the $s^{th}$ element equal to zero, the $s^{th}$ element is 1, $\langle s|$ denotes the corresponding row vector. Similarly $I_t$ is a diagonal matrix with all diagonal elements equal to 1 except the $tt^{th}$ elements which are equal to $-1$. $I_t$ can be written in the form $I_t = I - \sum_t 2|t\rangle\langle t|$. $U$ denotes an arbitrary unitary matrix and $U^{-1}$ is its inverse. The following results hold for arbitrary $U$, section 6 describes how to choose $U$ based on the specified probability distribution.

In this notation:

$$Q|s\rangle \equiv -(I_s U^{-1} I_t U)|s\rangle$$

$$= -(I - 2|s\rangle\langle s|)U^{-1}\left(I - \sum_t 2|t\rangle\langle t|\right)U|s\rangle$$

$$= -|s\rangle + 2|s\rangle + \sum_t 2\langle t|U|s\rangle U^{-1}|t\rangle$$

$$-\sum_t 4|s\rangle\langle s|U^{-1}|t\rangle\langle t|U|s\rangle$$

Note that $\langle t|U|s\rangle \equiv U_{ts}$. Also, since $U$ is unitary, $U^{-1}$ is equal to the transpose of the complex conjugate of $U$, and therefore $\langle s|U^{-1}|t\rangle = \langle t|U^*|s\rangle$ which is equal to $U_{ts}^*$. Therefore the above equation becomes:

$$Q|s\rangle = |s\rangle + \sum_t (2U_{ts})U^{-1}|t\rangle - \sum_t 4|U_{ts}|^2|s\rangle$$

Similarly, for any $t$ state:

$$QU^{-1}|t\rangle \equiv -(I_s U^{-1} I_t U)U^{-1}|t\rangle$$

$$= -I_s U^{-1} I_t |t\rangle$$

$$= (I - 2|s\rangle\langle s|)U^{-1}|t\rangle$$

$$= U^{-1}|t\rangle - 2U_{ts}^*|s\rangle$$

The analysis so far shows that, if there are $\eta$ $t$ states, then the $(\eta + 1)$ dimensional space defined by the vector $|s\rangle$ and the $\eta$ vectors $U^{-1}|t\rangle$ is preserved by the operator $Q$. The next two sections show that there is indeed a simpler two dimensional subspace that is also preserved by $Q$.

## 4. Single solution case

First of all consider the situation with a single $t$ state. This is the same as the generalized search situation considered previously [10]. In this section, we quickly rederive the main result since, in the next section, this will help to analyze the multi-solution case. The following two equations follow from the previous section:

$$Q|s\rangle = |s\rangle(1 - 4|U_{ts}|^2) + 2U_{ts}U^{-1}|t\rangle$$

$$QU^{-1}|t\rangle = U^{-1}|t\rangle - 2U_{ts}^*|s\rangle$$

Thus $Q$ is a transformation in the 2-dimensional vector space defined by the two vectors $|s\rangle$ and $U^{-1}|t\rangle$. In matrix notation this may be written as the following transformation:

$$Q\begin{bmatrix}|s\rangle \\ U^{-1}|t\rangle\end{bmatrix} = \begin{bmatrix}(1 - 4|U_{ts}|^2) & -2U_{ts}^* \\ 2U_{ts} & 1\end{bmatrix}\begin{bmatrix}|s\rangle \\ U^{-1}|t\rangle\end{bmatrix}.$$

In order to find the effect of repeated applications of $Q$, we use standard matrix analysis which consists of finding the eigenvalues and eigenvectors of the transformation matrix. Assuming $U_{ts}$ to be real and small, the two eigenvalues and eigenvectors are approximately:

$$\lambda_1 = 1 + 2iU_{ts}, v_1 = \begin{bmatrix}1 \\ -i\end{bmatrix} \&$$

$$\lambda_2 = 1 - 2iU_{ts}, v_2 = \begin{bmatrix}1 \\ i\end{bmatrix}.$$

The initial state vector is $|s\rangle$, which in terms of the eigenvectors may be written as $\frac{1}{2}(v_1 + v_2)$. After $\eta$ applications of $Q$, this transforms into:

$$\frac{1}{2}(v_1\lambda_1^\eta + v_2\lambda_2^\eta)$$ which may be simplified to:

$$\begin{bmatrix}\cos(2U_{ts}\eta) \\ \sin(2U_{ts}\eta)\end{bmatrix}.$$ Therefore, $\frac{\pi}{4U_{ts}}$ applications of $Q$, transform $|s\rangle$ into $U^{-1}|t\rangle$. From this, $|t\rangle$ may be obtained by a single application of $U$. After this, a measurement will reveal the target state, $t$, with certainty.

## 5. Multi-solution case

The analysis of section 4 breaks down if there is more than one $t$ state with different values of $U_{ts}$. This section extends the previous results. The relevant equations from section 3 are:

$$Q|s\rangle = |s\rangle\left(1 - 4\sum_t|U_{ts}|^2\right) + 2\sum_t U_{ts}U^{-1}|t\rangle$$

$$QU^{-1}|t\rangle = U^{-1}|t\rangle - 2U_{ts}^*|s\rangle \text{ (for each } t \text{ state.)}$$

Multiply the second equation by $U_{ts}$ and sum over all $t$ states. The equations now become -

$$Q|s\rangle = |s\rangle\left(1 - 4\sum_t|U_{ts}|^2\right) + 2\sum_t U_{ts}U^{-1}|t\rangle \text{ (as before)}$$

$$Q\sum_t U_{ts}U^{-1}|t\rangle = \sum_t U_{ts}U^{-1}|t\rangle - 2\sum_t|U_{ts}|^2|s\rangle$$

$Q$ is thus a transformation in the two dimensional com-

plex Hilbert space defined by $|s\rangle$ and $\sum_t U_{ts} U^{-1} |t\rangle$.

Normalizing the vector $\sum_t U_{ts} U^{-1} |t\rangle$ and denoting $\sqrt{\sum_t |U_{ts}|^2}$ by $u$, the above transformation can be represented as:

$$Q \begin{bmatrix} |s\rangle \\ \frac{1}{u}\sum_t U_{ts} U^{-1} |t\rangle \end{bmatrix} = \begin{bmatrix} (1-4u^2) & -2u \\ 2u & 1 \end{bmatrix} \begin{bmatrix} |s\rangle \\ \frac{1}{u}\sum_t U_{ts} U^{-1} |t\rangle \end{bmatrix}$$

This is exactly the same transformation as in the previous section with a single $t$ state. Therefore, as in the previous section, if $u \ll 1$, then $\frac{\pi}{4u}$ applications of $Q$ transform the vector $|s\rangle$ into $\frac{1}{u}\sum_t U_{ts} U^{-1} |t\rangle$. The case when $u$ is $O(1)$ is similar but not particularly interesting since this can be solved with a high probability by just applying $U$ to $|s\rangle$ once and carrying out an observation.

Note that the algorithm in the previous section was only shown to work when $U_{ts}$ was real. This section shows that a very similar result holds in the single solution case, even when $U_{ts}$ is complex.

**6. State vector engineering** Suppose that we are required to synthesize a specified superposition. The amplitude in each of $N$ states, denoted by $\bar{x}$, is required to be proportional to a given complex valued function $f(\bar{x})$. Since $f(\bar{x})$ can be scaled by an arbitrary factor, we assume, without loss of generality that the maximum value of $|f(\bar{x})|$ is less than or equal to $1$.

The following algorithm synthesizes the specified superposition in approximately $\frac{\pi}{4} \sqrt{\frac{N}{\sum_{\bar{x}} |f(\bar{x})|^2}}$ steps.

One immediate application of this algorithm is in sampling according to an arbitrary probability distribution in $O(\sqrt{N})$ steps. For this a quantum superposition is generated and then through a measurement in an appropriate basis, a sample is obtained.

*Solution* Define $N \equiv 2^n$ states by $n$ qubits, denoted by $\bar{x}$. Include an additional ancilla qubit. Initialize the state so that all qubits are in the $0$ state - the state of the whole system is denoted by: $(0, \bar{0})$ (the first $0$ denotes a single qubit in the $0$ state and $\bar{0}$ denotes each of the other $n$ qubits in the $0$ state.)

Next consider the following unitary operations (which constitute the building blocks for our algorithm):

$U_1$: leave the first qubit unaltered and apply a W-H transform to the other $n$ qubits.

$U_2$: carry out a conditional rotation of the first qubit so that the state: $(0, \bar{x})$ gets transformed into the following superposition:

$(f(\bar{x})(0, \bar{x}) + \sqrt{1-|f(\bar{x})|^2}(1, \bar{x}))$, i.e. the amplitude of the state $(0, \bar{x})$ is $f(\bar{x})$ and the amplitude of the state $(1, \bar{x})$ is $\sqrt{1-|f(\bar{x})|^2}$. This type of unitary operation has previously been used in quantum computing algorithms, e.g. in the mean estimation algorithm in [10]. It can be accomplished by first transferring $f(\bar{x})$ into the phase through conditional phase inversion and then converting it into amplitude information.

$I_t$: in case the first qubit is $0$, invert the phase; if the first qubit is $1$, leave it unchanged. In other words, states with the first qubit in the $0$ state are $t$ states.

$I_s$: in case all the qubits (including the first qubit) are $0$, invert the phase; else leave it unchanged, i.e. $(0, \bar{0})$ is the $s$ state.

Clearly if we start with the $(0, \bar{0})$ state and apply $U_1$ and then $U_2$, the amplitude in the $(0, \bar{x})$ state is $\frac{f(\bar{x})}{\sqrt{N}}$. In other words, if we define the composite operation $U \equiv U_2 U_1$, the $s$ state as $(0, \bar{0})$ and the $t$ states as the $(0, \bar{x})$ states; then $U_{ts}$, the matrix element between $s$ and the relevant $t$ state is $\frac{f(\bar{x})}{\sqrt{N}}$.

It immediately follows from section 5, that by starting with the $(0, \bar{0})$ state, and applying the sequence of operations defined by $Q \equiv -(I_s U^{-1} I_t U)$, $\frac{\pi}{4}\sqrt{\frac{N}{\sum_{\bar{x}}|f(\bar{x})|^2}}$ times, followed by a single application of $U$, we get the first qubit in the $0$ state and the remaining $n$ qubits in a superposition with the amplitude of the $\bar{x}$ state as $\frac{f(\bar{x})}{\sqrt{\sum_{\bar{x}}|f(\bar{x})|^2}}$.

## 7. Observations

(i) Let $f(\bar{x})$ be 1 at $\eta$ points in the domain and zero everywhere else. Since $\sum_{\bar{x}}|f(\bar{x})|^2 = \eta$, the algorithm needs $\frac{\pi}{4}\sqrt{\frac{N}{\eta}}$ steps. After this it reaches the same superposition as reached in the basic quantum search algorithm with $\eta$ solutions and it needs exactly the same number of steps as the quantum search algorithm to reach this.

(ii) The number of steps required by the algorithm depends on $\sum_{\bar{x}}|f(\bar{x})|^2$. In case this quantity is not known in advance, the superposition can still be synthesized in $O(\sqrt{N})$ steps, by trying out the algorithm with a few carefully chosen runtimes. After this, the ancilla qubit is measured - the algorithm is repeated until this is observed to be $0$ (it can be shown that, with appropriately chosen runtimes, the probability of not getting even a single $0$ falls exponentially with the square of the number of times the procedure is repeated [7]). Once a $0$ is observed, the remaining $n$ qubits immediately collapse into the desired superposition.

(iii) The algorithm assumed that the function $f(\bar{x})$ was normalized so that its maximum value was equal to $1$. The algorithm as presented in this section is equally valid for different $f(\bar{x})$, provided the maximum value does not exceed $1$. For example, if the desired probability at each $\bar{x}$ is specified (i.e. $|f(\bar{x})|^2$), the value of $\sum_{\bar{x}}|f(\bar{x})|^2$ becomes 1 and the algorithm needs exactly $\frac{\pi}{4}\sqrt{N}$ iterations to attain this distribution. In general, the number of steps required is smaller if we can choose a larger constant to scale $f(\bar{x})$. This is maximized for the choice made in this section where the maximum value of $f(\bar{x})$ is 1.

(iv) The analysis presented above is very similar for the situation where the initial state $|s\rangle$, instead of being a basis state is an arbitrary superposition as considered in [3]. The main difference is that in the computational basis, $I_s$ (the matrix that inverts the phase of the source state) will not be diagonal. Of course, it will be diagonal in any basis that includes $|s\rangle$. In terms of the Dirac notation it will still be described as $I_s = I - 2|s\rangle\langle s|$, i.e. it reflects the $|s\rangle$ state and leaves all orthogonal states unchanged.

(v) A superposition with multiple target states, of the type discussed in section 5, is a pure state in a different basis. Thus the algorithm rotates from a given source state to another pure state. This is just like quantum search except that the final state is a basis state in a different basis - the quantum search algorithm could only rotate into a pure basis state in the same basis.

**8. Conclusion** The quantum search algorithm is perhaps the simplest possible quantum mechanical algorithm that yields a significant advantage over a classical algorithm. In addition to the basic result that a quantum computer can search a domain of size $N$ in $O(\sqrt{N})$ steps, its contribution has been to inspire several new ideas and algorithms. This paper presents the most recent such development which shows that the quantum search class of algorithms can be made to work in the presence of multiple target states. Such a result may seem obvious for classical computers; however, for quantum computers one needs to take into account interference effects due to the different solutions and the situation is more complicated.

An application has been presented where one can generate a sample according to an arbitrary probability distribution in a number of steps which is only a square-root of that required by a classical algorithm. Another immediate application is in extending the framework of the quantum search class of algorithms with arbitrary unitary transformations of the type discussed in [5] & [10], so that they are no longer limited to problems with a single solution.

Just like previous versions of the quantum search algorithm, the framework here is completely general and it is expected that it will find more applications.

**9. Acknowledgements** Thanks to Ashwin Nayak, Norm Margolus, Hein Roehrig and Charles Bennett for going through the paper and making valuable comments.